\documentclass{article}

\usepackage[latin1]{inputenc} 
\usepackage{url}
\usepackage{hyperref}
\usepackage[scale=0.7]{geometry}
\usepackage{setspace}

\newcommand{\oper}[1]{{\sc#1}}
\newcommand{\se}{\Rightarrow}
\newcommand{\sse}{\Leftrightarrow}

\newcommand{\mytitle}{On the Coordinator's Rule for Fast Paxos}

\hypersetup{
     pdftitle={\mytitle}, 
     pdfauthor={Gustavo M.  D. Vieira, Luiz E.  Buzato},
     pdfdisplaydoctitle=true,
     hidelinks
}

\title{\mytitle}

\author{Gustavo M.   D.  Vieira\thanks{Financially supported by CNPq,
    under grant 142638/2005-6.}  \and Luiz E.  Buzato}

\date{ \small Institute of Computing---UNICAMP \\
       \small Caixa Postal 6176 \\
       \small 13083-970 Campinas, São Paulo, Brasil \\
       \small \url{{gdvieira, buzato}@ic.unicamp.br}}

\begin{document}

\maketitle


\begin{abstract}
  Fast Paxos is an algorithm  for consensus that works by a succession
  of rounds,  where each  round tries  to decide a  value $v$  that is
  consistent  with   all  past  rounds.   Rounds  are   started  by  a
  coordinator process  and consistency is guaranteed by  the rule used
  by this  process for the selection  of $v$ and by  the properties of
  process sets  called quorums. We  show a simplified version  of this
  rule  for the specific  case where  the quorums  are defined  by the
  cardinality of these process sets.  This rule is of special interest
  for implementors of the algorithm.
\end{abstract}

\section{Introduction}

The problem of deciding a single value out of a set of values proposed
by processes  is known as the \emph{consensus}  problem.  This problem
is easy to  solve in the absence of failures, but  it is impossible to
solve in an  asynchronous distributed system even if  a single process
fails    by   permanently    stopping~\cite{fischer85}.     A   better
approximation  of the failures  that processes  of a  real distributed
system  can suffer  is  the one  where  processes stop  but may  later
recover.   Unfortunately,  the  impossibility  also  holds  for  these
systems.  One of the ways to get around the impossibility is to design
algorithms  that do not  violate their  safety requirements  while the
system behaves asynchronously and are  certain to make progress if the
system  behaves partially  synchronously  for periods  long enough  to
satisfy the progress requirements.  Designing consensus algorithms for
the asynchronous crash-recovery model is a difficult task of practical
interest     and    probably    Paxos~\cite{lamport98}     and    Fast
Paxos~\cite{lamport06a} are the most studied solutions so far.

Fast Paxos solves the consensus problem through a succession of rounds
that  lead to  the choice  of  the consensus  value. In  each round  a
distinguished  process,  the  \emph{coordinator}, is  responsible  for
picking  a single  value using  a  rule that  is based  on quorums  of
processes. Quorums of successive rounds  are used to guarantee that if
a single  value has been  chosen or might  ever be chosen  in previous
rounds  then the  same value  is  going to  be chosen  in the  current
round. Thus, quorums are fundamental  to the correctness of Fast Paxos
because they are ultimately responsible for the validity of consensus.

Lamport~\cite{lamport06a} shows how quorums can be characterized using
the cardinality of sets of  processes, defining what minimum number of
processes represents a quorum.   However, he defines the coordinator's
rule  in  terms of  quorum  sets  and  general set  operations.   This
complete characterization of the coordinator's rule is perfect for the
purposes of his work, but it  does not address thoroughly the needs of
a  programmer  who  wants   to  implement  it.   Therefore,  the  main
contributions  of this  paper are  (i) an  interpretation of  the Fast
Paxos coordinator's rule only in  terms of the cardinalities of quorum
sets, and  (ii) its simplification.  The  simplified interpretation is
efficient, easier  to implement  and test; it  can help  developers to
create reliable implementations of  Fast Paxos.  This is important, as
the use of Fast Paxos to build fault-tolerant applications is bound to
require the execution of a very large number of consensus instances.

\subsection{Fast Paxos}

Before detailing the  coordinator's rule, it is useful  to give a very
brief  overview of Fast  Paxos; a  complete description  of it  can be
found  in~\cite{lamport06a}.  The  algorithm is  easier to  explain in
terms of  reactive agents  that represent a  role, such that  a single
process can  enact multiple  agents, with each  one of them  playing a
different role.  An agent can enact one of the following main roles: a
\emph{proposer} that can propose  values by sending them to acceptors,
an \emph{acceptor}  that chooses a single value,  or a \emph{learner}
that learn what value has been chosen.

To solve  consensus, Fast Paxos  agents execute multiple  rounds, each
round has  a coordinator and  may be either  a \emph{fast} round  or a
\emph{classic} round.  Positive integers are used to uniquely identify
rounds, each  identifier determines the coordinator  and indicates the
round  type: fast  or classic.   Regardless  of its  type, each  round
progresses through two phases with two steps each:
\begin{itemize}
\item In  Phase 1a  the coordinator sends  a message  requesting every
  acceptor  to participate  in  round $i$.   An  acceptor accepts  the
  invitation if  it has not  already accepted to participate  in round
  $j \geq i$, otherwise  it declines the invitation by  simply ignoring it.
\item  In Phase  1b every  acceptor that  has accepted  the invitation
  answers  to the  coordinator with  a reply  that contains  the round
  number and the  value of the last  vote it has cast for  a value, or
  \textsl{null} if it has not voted. 
\item  In Phase  2a,  if the  coordinator  of round  $i$ has  received
  answers from a quorum of acceptors  then it executes its rule on the
  set of values suggested by acceptors  in Phase 1b and picks a single
  value $v$.   It then asks  the acceptors to  cast a vote for  $v$ in
  round $i$, if  $v$ is not \textsl{null}, otherwise,  if the round is
  fast  the coordinator sends  a \emph{any}  message to  the proposers
  indicating that any value can be chosen in round $i$.  In this case,
  the proposers  ask the acceptors to cast  a vote for a  value $v$ of
  their choice in round $i$.
\item In Phase  2b, after receiving a request to cast  a vote from the
  coordinator or from one of  the proposers, acceptors can either cast
  a vote  for $v$ in round  $i$, if they  have not voted in  any round
  $j \geq i$, otherwise, they ignore the  vote request.  Votes are cast by
  sending them together with the round identifier to the learners.
\item Finally, a  learner learns that a value $v$  has been chosen if,
  for some round  $i$, it receives Phase 2b messages  from a quorum of
  acceptors announcing that they have all voted for $v$ in round $i$.
\end{itemize}

As Fast Paxos agents may crash and recover, they must save their state
in  stable memory  so that  agents, once  recovered, can  remember the
votes they have  cast earlier.  The sequence of  steps described above
imply that  a learner can  only learn the  value of consensus  after a
period of  at least  four message delays.   If numerous  executions of
Fast Paxos are required, then it  is possible to run Phase 1 and Phase
2a only once for all these instances.  This factorization of phases is
carried out immediately after the  election of a coordinator.  At this
point,  most of  the consensus  instances have  not been  started yet,
allowing the  coordinator to send  Phase 2a \emph{any}  messages.  The
improvement brought  about by  this factorization allows  consensus in
two   message  delays,   making  Fast   Paxos  an   optimal  consensus
algorithm~\cite{lamport06b}.  Unfortunately,  Fast Paxos cannot always
be fast.  Proposers can  propose two different values concurrently, in
this   case,  their   proposals  may   collide.   Also,   process  and
communication failures  may block a round  from succeeding.  Different
recovery  mechanisms can be  implemented to  deal with  collisions and
failures, but eventually the coordinator intervention may be necessary
to start  a new round~\cite{lamport06a}.   Any process can act  as the
coordinator as  long as it follows  the rule for choosing  a value, if
any, that is proposed in Phase 2a.

As already mentioned, quorums  are fundamental for Fast Paxos. Quorums
are set  of processes and each  round has a set  of quorums associated
with it,  classic quorums for a  classic round and fast  quorums for a
fast round.  For  the proper operation of the  algorithm, quorums have
to  satisfy  properties on  the  sets  of  processes that  form  them.
Specifically, any  two quorum sets must have  a non-empty intersection
and any quorum and any two  fast quorums from the same round must also
have a non-empty  intersection~\cite{lamport06a}.  There are many ways
to define quorums,  but a very interesting one from  the point of view
of process fault tolerance is  the definition based only on the number of
processes contained  in each quorum.   The definition of  quorum using
this    parameter   is   straightforward    and   is    described   in
Section~\ref{cardinality}. 

The coordinator's  rule determines how a  coordinator can consistently
start a new round,  after collecting information about previous rounds
from the  acceptors. That  is, for each  round $i$ the  coordinator is
about to start, it must know if  a value $v$ had been decided or might
have been decided in previous rounds $j<i$.  The coordinator's rule of
Fast Paxos must take into account  that in a fast round, more than one
value might  have been proposed  and voted concurrently.   Quorums are
defined  to  guarantee that  only  one,  if  any, of  the  conflicting
proposed   values  is   selected  through   the  application   of  the
coordinator's   rule.   Section~\ref{rule1}   presents   the  original
coordinator's rule as defined  in~\cite{lamport06a} and then shows how
this  rule  can  be  effectively implemented  using a cardinality-based
definition of quorums. Section~\ref{rule2} brings our derivation of a
simplified cardinality-based coordinator rule, it is stricter than the
one presented  in Section~\ref{rule1}, but it is  easier to understand
and to  implement. Section~\ref{conclu} closes the  work by commenting
on the practical value of  our main result: a simplified coordinator's
rule for Fast Paxos.

\section{Choosing Quorums}
\label{cardinality}

The  quorum requirements  for  Fast  Paxos assert  that:  (a) any  two
quorums must have non-empty intersection, (b) any two fast quorums and
any  classic or  fast  quorum from  the  same round  have a  non-empty
intersection~\cite{lamport06a}.   We can  satisfy these  conditions by
considering only  the number of process  in each quorum,  where $N$ is
the number  of acceptors, and  $F$ and $E$  are the maximum  number of
failed     acceptors      in     classic     and      fast     rounds,
respectively~\cite{lamport06a}.  A classic quorum is formed by $N - F$
acceptors  and  $N  -  E$  acceptors  form  a  fast  quorum.   As  the
requirements  for fast  quorums  are always  stricter  than those  for
classic quorums,  we can  always assume that  $E \leq F$.   The quorum
conditions~\cite{lamport06a} are then stated as:

\begin{eqnarray}
N & > & 2F \label{eq-classic} \\
N & > & 2E + F \label{eq-fast}
\end{eqnarray}

For a fixed  $N$, $F$ and $E$ can be chosen  in various different ways
and  a natural  way  of choosing  them  is by  maximizing  one or  the
other~\cite{lamport06a}.  As we have  $E \leq F$, maximizing $E$ leads
to $E = F$. Thus, we can satisfy the system only with $N > 3F$ and:
\[
N > 3F \sse F < N/3 \sse F \leq \lceil N/3 \rceil - 1
\]
For this  case, the cardinality  of any classic quorum  ($|Q_{c}|$) or
fast quorum ($|Q_{f}|$), expressed only as a function of $N$, is:
\[
|Q_{c}| = |Q_{f}| \geq N - \lceil N/3 \rceil + 1 
                  \geq \lfloor 2N/3 \rfloor + 1
\]

If instead  we maximize $F$, the limit  for its value is  given by the
Equation~\ref{eq-classic}, thus:
\[
N > 2F \sse F < N/2 \sse F \leq \lceil N/2 \rceil - 1
\]
In this  case $E$ must be chosen  to satisfy Equation~\ref{eq-fast},
considering the value of $F$ we have just chosen:
\[
N > 2E + F \sse N > 2E + \lceil N/2 \rceil - 1 
           \sse 2E \leq N - \lceil N/2 \rceil 
           \sse E \leq \lfloor N/4 \rfloor 
\]
For this case,  the cardinality of any classic  quorum ($|Q_{c}|$) and
fast quorum ($|Q_{f}|$), expressed only as a function of $N$, is:
\begin{eqnarray*}
|Q_{c}| \geq  N - \lceil N/2 \rceil + 1 
        \geq \lfloor N/2 \rfloor + 1 \\
|Q_{f}| \geq  N - \lfloor N/4 \rfloor 
        \geq \lceil 3N/4 \rceil 
\end{eqnarray*}

\section{Coordinator's Rule}
\label{rule1}

The original coordinator's rule for Fast Paxos~\cite{lamport06a} is:

\begin{singlespace}
\begin{tabbing}
\oper{let} \= $Q$ \= be any $i$-quorum of acceptors that have reported
              their last votes to the \\
           \> \> coordinator. \\
           \> $vr(a)$  and $vv(a)$ be the round  and the  value voted
              by acceptor $a$. \\
           \> $k$ \> be the largest value of $vr(a)$ for all $a \in Q$.\\
           \> $V$ \> be the set of values $vv(a)$ for all $a \in Q$ with
              $vr(a) = k$. \\
           \> $O4(v)$ be true iff there is a $k$-quorum $R$ such that 
              $vr(a) = k$ and $vv(a) = v$ \\
           \> \> for all $a \in (Q \cap R)$.
\end{tabbing}
\begin{tabbing}
\oper{if} $k = 0$ \= \oper{then} let $v$ be any proposed value. \\
                  \>  \oper{else} \= \oper{if}  $V$ contains  a single
                  element \\
                  \> \> \oper{then} let $v$ equal that element. \\
                  \> \> \oper{else} \= \oper{if} there is some $w \in V$
                  satisfying $O4(w)$ \\
                  \> \> \> \oper{then} let $v$ equal that $w$ (unique). \\
                  \> \> \> \oper{else} let $v$ be any proposed value.
\end{tabbing}
\end{singlespace}

We  now  show  how this  rule  can  be  interpreted  in terms  of  the
cardinality-based  quorum   definitions  presented  in   the  previous
section. When  $k = 0$  or $V$ contains  a single element the  rule is
trivial  to  evaluate  no   matter  the  quorum  implementation  used.
However, the evaluation of $O4(w)$ is more complex because it requires
the  evaluation of  all  possible  intersections $Q  \cap  R$ for  all
$k$-quorums  $R$.  Considering  only  the cardinality  of the  quorums
involved  we have  that  $O4(w)$ is  true  if at  least  $|Q \cap  R|$
acceptors voted  for $w$ for  some $R$.  As  we don't know,  and don't
want to know, all possible  quorums $R$, we must consider the smallest
possible  $|Q  \cap R|$,  assuming  as  implied  by $O4(w)$  that  all
acceptors  \emph{outside} of  $Q$ also  voted for  $w$ in  ballot $k$.
Considering that $V$ can only contain more than one element if $k$ was
a fast round,  we have two situations: $i$ is a  classic quorum or $i$
is a fast quorum.  Let $T$ be the number of votes for the value $w$ in
$V$.  If we want $T$ to be  at least as large as the smallest $|Q \cap
R|$ then we have:
\[
T \geq \left\{
  \begin{array}{ll}
    N - E - F & \mbox{if $i$ is classic} \\
    N - 2E    & \mbox{if $i$ is fast} \\      
  \end{array}\right. 
\]

$O4(w)$ can  now be evaluated by  simply counting the  number of votes
for $w$  in  $V$.  So, any  value $w$ that  satisfies the
condition  above  satisfies  $O4(w)$  and  is  by  definition  unique.
Considering that  $E < (N -
F) / 2$  from Equation~\ref{eq-classic}, when $i$ is a
classic round, we have:
\[
T \geq N - E - F \se T > N - (N - F) / 2 - F 
        \sse T > (N - F)/2 \sse T \geq \lfloor |Q_{c}|/2 \rfloor +1
\]
Similarly, for the case where $i$ is a fast round we have:
\begin{eqnarray*}
T \geq N - 2E &\se& T > N - E - (N - F) / 2 \sse 2T > N - E + (F - E) \\
              &\se& T > (N - E)/2 \sse T \geq \lfloor |Q_{f}|/2 \rfloor +1
\end{eqnarray*}

In all cases $T$ is at least  as large as $\lfloor |Q|/2 \rfloor + 1$,
so if any value $w$ satisfies $O4(w)$, then it has been voted in round
$k$ by a majority of processes \emph{inside} the quorum $Q$.

\section{Simplified Coordinator's Rule}
\label{rule2}

We have  shown that a value  $w$ satisfies $O4(w)$ if  this value have
been voted in round $k$ by a majority of acceptors in $Q$.  We can use
this observation  to derive a  simplified coordinator's rule  for Fast
Paxos.  First, the  fact that $w$ has been voted by  a majority in $Q$
implies that  it is the  value most often  voted in $V$. Thus,  we can
check  this   condition before  testing  if $w$  satisfies
$O4(w)$, obtaining an equivalent coordinator's rule.

\begin{singlespace}
\begin{tabbing}
\oper{if} $k = 0$ \= \oper{then} let $v$ be any proposed value. \\
                  \>  \oper{else} \= \oper{if}  $V$ contains  a single
                  element \\
                  \> \> \oper{then} let $v$ equal that element. \\
                  \> \> \oper{else} \= \oper{if} there is a single $w \in
                  V$ voted most often \\
                  \> \> \> \oper{then} \= \oper{if} $w$ satisfies $O4(w)$ \\
                  \> \> \> \> \oper{then} let $v$ equal that $w$. \\
                  \> \> \> \> \oper{else} let $v$ be any proposed value. \\
                  \> \> \> \oper{else} let $v$ be any proposed value.
\end{tabbing}
\end{singlespace}

If $w$  do not  satisfy $O4(w)$, we  are free  to choose any  value as
$v$. We  use this freedom to  always select the most  often voted $w$.
We have now removed some  freedom from the coordinator, but all values
$w$ that satisfy  $O4(w)$ are correctly selected.  We can remove
the $O4(w)$ test, obtaining the following rule:

\begin{singlespace}
\begin{tabbing}
\oper{if} $k = 0$ \= \oper{then} let $v$ be any proposed value. \\
                  \>  \oper{else} \= \oper{if}  $V$ contains  a single
                  element \\
                  \> \> \oper{then} let $v$ equal that element. \\
                  \> \> \oper{else} \= \oper{if} there is a single $w \in
                  V$ voted most often \\
                  \> \> \> \oper{then} let $v$ equal that $w$. \\
                  \> \> \> \oper{else} let $v$ be any proposed value.
\end{tabbing}
\end{singlespace}

If $V$  contains a single element,  then this element  surely has been
voted most  often than any other  element in $V$. Thus,  we can remove
the single element test, giving our final simplified rule:

\begin{singlespace}
\begin{tabbing}
\oper{if} $k = 0$ \= \oper{then} let $v$ be any proposed value. \\
                  \> \oper{else} \= \oper{if} there is a single $w \in
                  V$ voted most often \\
                  \> \> \oper{then} let $v$ equal that $w$. \\
                  \> \> \oper{else} let $v$ be any proposed value.
\end{tabbing}
\end{singlespace}

\section{Conclusion}
\label{conclu}

We    have   showed    how    the   coordinator's    rule   of    Fast
Paxos~\cite{lamport06a} can be  simplified by resorting exclusively to
counting the  number of  votes for each  of the proposed  values.  Our
rule is more restrictive than the original rule, as for some consensus
instances it forbids the coordinator of freely choosing any value when
he  would be  allowed otherwise  by the  original rule.   However, the
restriction  imposed  by  the  simplification  does not  lead  to  any
disadvantage because if some value  received votes in a previous round
and  the  consensus value  isn't  decided  yet,  it is  reasonable  to
consider that the coordinator will  try to decide on that value first.
Our simplified rule is easier  to implement, has the advantage that it
is independent of the type of a round (fast or classic), and it has to
consider only the cardinality of the quorum $Q$.

\bibliographystyle{plain}
\bibliography{ipl-paxos}

\begin{thebibliography}{1}

\bibitem{fischer85}
Michael~J. Fischer, Nancy~A. Lynch, and Michael~S. Paterson.
\newblock Impossibility of distributed consensus with one faulty process.
\newblock {\em J. ACM}, 32(2):374--382, 1985.

\bibitem{lamport98}
Leslie Lamport.
\newblock The part-time parliament.
\newblock {\em ACM Trans. Comput. Syst.}, 16(2):133--169, 1998.

\bibitem{lamport06a}
Leslie Lamport.
\newblock Fast {P}axos.
\newblock {\em Distrib. Comput.}, 19(2):79--103, October 2006.

\bibitem{lamport06b}
Leslie Lamport.
\newblock Lower bounds for asynchronous consensus.
\newblock {\em Distributed Computing}, 19(2):104---125, June 2006.

\end{thebibliography}

\end{document}